\documentclass[prl, twocolumn, a4paper, superscriptaddress, showpacs, amsmath, amssymb, footinbib, floatfix]{revtex4}

\usepackage[utf8]{inputenc}
\usepackage[english]{babel}
\usepackage{graphicx}

\newcommand{\rmd}{\mathrm d}

\newcommand{\BM}[1]{\mbox{\boldmath$#1$}}
\newcommand{\Bm}[1]{\mbox{\scriptsize\boldmath$#1$}}

\begin{document}

\title{Submicron plasticity: yield stress, dislocation avalanches, and velocity distribution}

\author{P\'eter Dus\'an Isp\'anovity}
\email{ispanovity@metal.elte.hu}
\affiliation{Department of Materials Physics, E\"otv\"os University Budapest,
H-1517 Budapest POB 32, Hungary}
\affiliation{Paul Scherrer Institut, CH-5232 Villigen PSI, Switzerland}

\author{Istv\'an Groma}
\affiliation{Department of Materials Physics, E\"otv\"os University Budapest,
H-1517 Budapest POB 32, Hungary}

\author{G\'eza Gy\"orgyi}
\affiliation{Department of Materials Physics, E\"otv\"os University Budapest,
H-1517 Budapest POB 32, Hungary}

\author{Ferenc F.~Csikor}
\affiliation{Department of Materials Physics, E\"otv\"os University Budapest,
H-1517 Budapest POB 32, Hungary}

\author{Daniel Weygand}
\affiliation{Karlsruhe Institute of Technology, izbs, Kaiserstra{\ss}e 12,
D-76131 Karlsruhe, Germany}

\begin{abstract}
The existence of a well defined yield stress, where a macroscopic piece of crystal
begins to plastically flow, has been one of the basic observations of materials
science. In contrast to macroscopic samples, in micro- and nanocrystals the
strain accumulates in distinct, unpredictable bursts, which makes controlled
plastic forming rather difficult. Here we study by simulation, in two and three
dimensions, plastic deformation of submicron objects under increasing stress. We
show that, while the stress-strain relation of individual samples exhibits
jumps, its average and mean deviation still specify a well-defined critical
stress, which we identify with the jamming-flowing transition. The statistical
background of this phenomenon is analyzed through the velocity distribution of
short dislocation segments, revealing a universal cubic decay and an
appearance of a shoulder due to dislocation avalanches. Our results can help to
understand the jamming-flowing transition exhibited by a series of various
physical systems.
\end{abstract}

\maketitle

Understanding the nature of irreversible plastic deformation is a crucially important issue in current research in materials sciences. It has been known for a long time that macroscopic size materials begin to yield at a certain stress level, depending on several material parameters. On macroscopic scales the flowing regime is traditionally described by constitutive laws, envisaging plasticity as a smooth and steady flow both in time and space. In the last decade, however, a completely new picture has emerged. By analyzing the emitted sound waves during deformation it was observed that plastic deformation is characterized by intermittent bursts of activity \cite{weiss_lahaie_grasso, carmen_nature, weiss_science}. Furthermore, recent compression tests carried out on Ni microcrystals \cite{uchic_science, dimiduk_science} revealed that, if the specimen size is in the range of several $\mu$m and below, discontinuous deformation results in steps of irregular spacing on the stress-strain curve. The related strain bursts correspond to the sudden, collective motion of line-like lattice defects, the dislocations. The size  distribution of the dislocation avalanches decreases by a universal power-law with exponent $1.5$. The distribution was recovered by experiments \cite{brinckmann_prl,zaiser_schwerdfeger}, by two- (2D) and three-dimensional (3D) discrete dislocation dynamics (DDD) simulations \cite{carmen_nature,csikor_science}, and by analytical modeling \cite{koslowski, zaiser_moretti}. These findings indicate that intermittent dislocation avalanches are an intrinsic feature of the plasticity of crystals, and their size distribution is not affected by the details of the deformation.

In nanoscale applications this behavior has two important consequences. Firstly,
the increased relative size of the fluctuations makes difficult to control the
plastic forming process \cite{csikor_science}. Secondly, at small specimen sizes
the yield stress is not well-defined any more. According to the conventional
definition, the yield stress of a specimen is the external stress at $0.2$\%
plastic strain. If the specimen size is below several $\mu$m, then, as a result
of strain fluctuations and statistical effects in the microstructure, this value
varies specimen by specimen \cite{brinckmann_prl} prohibiting a
material-specific yield stress definition. One can thus raise the question,
whether it is possible to give a new physical definition for the yield stress,
or in other words, whether the strength of a micron size specimen can be defined
at all.

Similar phenomena are observed in completely different physical systems, too.
Granular materials, such as sand piles, also exhibit yield stress, and at higher
external driving forces deformation occurs in distinct
avalanches \cite{granular_matter_review}. Plate
tectonics \cite{gutenberg_richter}, fracture dynamics \cite{zapperi_nature},
vortex lattices in superconducting films \cite{blatter_vortex_review}, and
dynamics of domains in ferromagnets \cite{durin_zapperi} are further examples of
such processes. The common characteristics of these systems are marginal
stability, power-law distributions without characteristic length- or timescales,
and driving forces that vary much slower than the internal relaxation
processes \cite{sammonds}. For such problems the term ``self-organized
criticality'' is widely used \cite{bak,sethna_nature}.

Although irregular plastic response of submicron crystalline materials is by now
conceived as a self-organized critical phenomenon \cite{carmen_andrade,
dimiduk_science,sammonds,csikor_science}, a physical understanding, that is, a
phenomenology for plastic flow based on statistical properties of strain
avalanches is still lacking. In this paper we present a statistical analysis of
the fluctuating stress-strain response of individual specimens, and of the
velocity distribution $P(v)$ of short dislocation segments. Our main proposition
is that the $P(v)$ holds the key to many empirical aspects of the flow. It
exhibits a specific transition at the onset of material yielding seen on the
average stress-strain characteristics.

Concerning the dynamics of dislocations, it is commonly assumed that the motion of dislocations is essentially overdamped. Thus the glide velocity is proportional to the glide component of the acting  Peach--Koehler  force \cite{hirth_lothe} per unit length $F_\text{g}$, that is, $v = B^{-1} F_\text{g}$, where $B$ is the drag coefficient of the dislocations \cite{hirth_lothe}.   Here $F_\text{g} = b \tau_\text{rs} + F_\text{s}$, where $b$ is the magnitude of the Burgers vector and $\tau_\text{rs}$ is the sum of the external stress and local resolved shear stress, generated by the other dislocation segments via long-range interactions, while $F_\text{s}$ denotes the short-range forces related to dislocation self-interaction and junction formation \cite{hirth_lothe}.

\begin{figure}[!ht]
\vspace*{-0.4cm}
\begin{center}
\vspace*{-3cm}
(a)
\begin{minipage}{3cm}
\vspace*{3.8cm}
 \includegraphics[angle=-0,width=3cm]{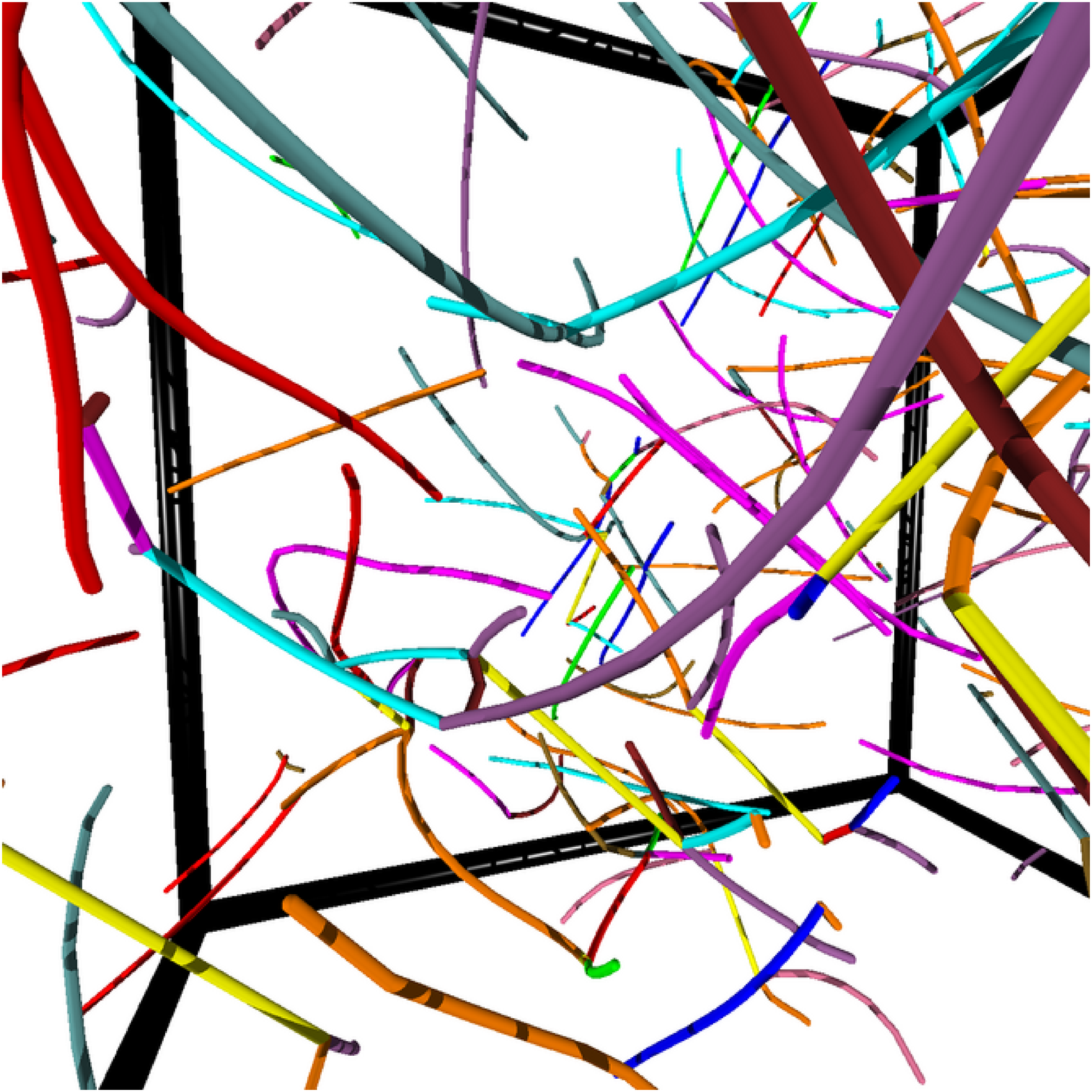}\hspace{15pt}
\end{minipage}
\hspace*{0.5cm}
(b)
\begin{minipage}{3.7cm}
\vspace*{3cm}
 \includegraphics[angle=-0,width=3.7cm]{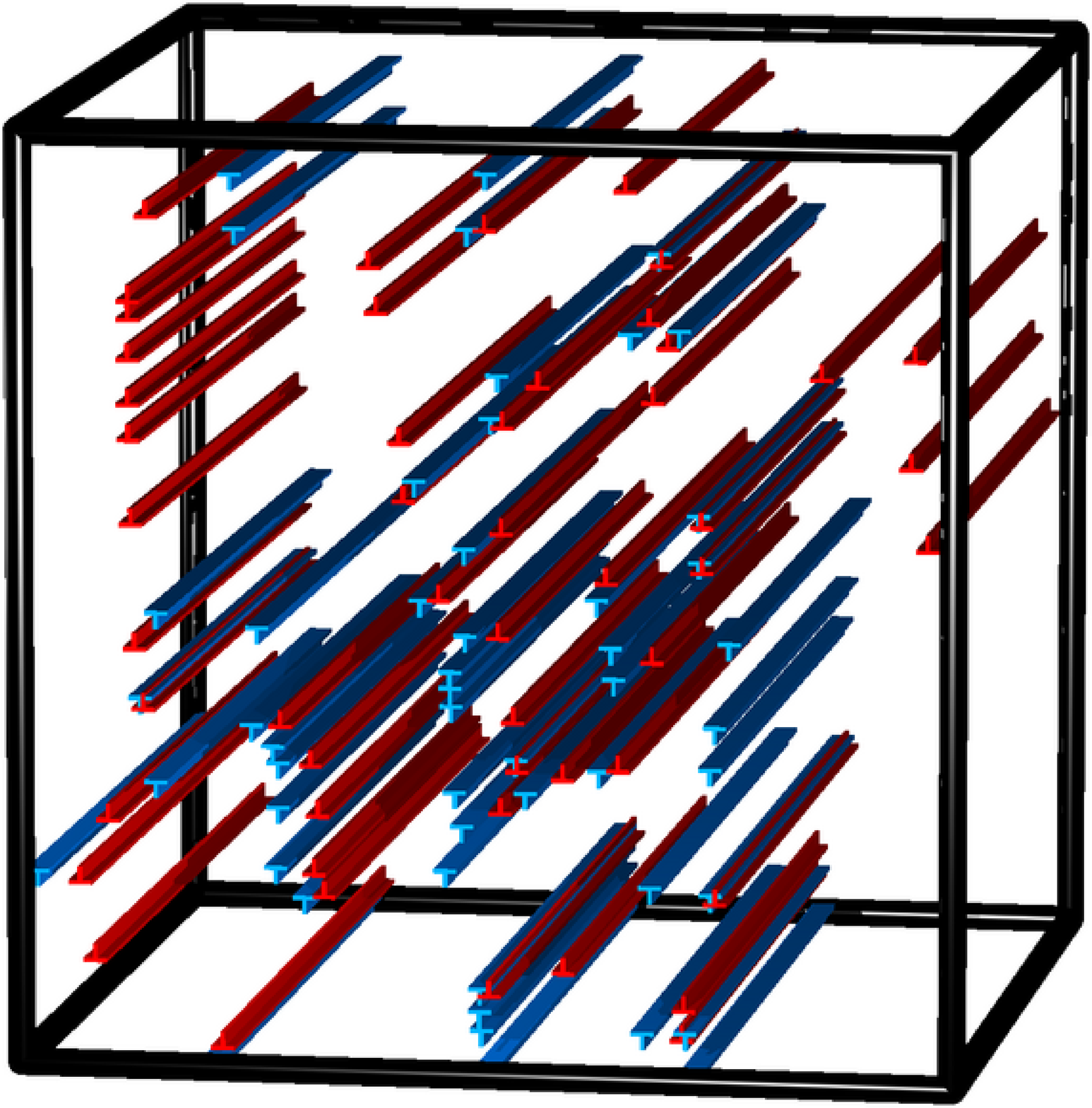}
\end{minipage}
\end{center}
\vspace*{-0.4cm}
\caption{Dislocation configurations obtained by 2D and 3D simulations.
(a) Snapshot of a three-dimensional simulation. Different
colors indicate dislocations on the 12 different slip systems.
(b) Snapshot of a two-dimensional system consisting of
straight parallel edge dislocations. \label{fig:1}}
\end{figure}

To study the submicron plastic response by DDD we simulated in 3D several
statistically equivalent realizations of the yielding of an $L = 0.46$ $\mu$m
edge size aluminium cube subject to compression with constant stress rate and
free side boundaries (for simulation details see \cite{weygand_friedman}).
The initial dislocation density was $8\times 10^{13}$ m$^{-2}$. Part
of a typical configuration is shown in Fig.~\ref{fig:1}(a) \footnote{See EPAPS Document No. [] for example simulation movies. For more information on EPAPS, see http://www.aip.org/pubservs/epaps.html.}. Motivated by the fact that in
2D the avalanche size distribution was found to be very similar to
the 3D case \cite{carmen_nature, zaiser_moretti} simulations on 2D systems
consisting of straight parallel edge dislocations, oriented for single slip with
periodic boundary conditions, were also performed. A typical 2D system is seen
in Fig.~\ref{fig:1}(b) \footnotemark[\value{footnote}]. Although several 3D ingredients, like dislocation multiplication,
junctions and forest dislocations are absent from the 2D system, there is ground
to expect that the essential feature governing strain avalanches are included in
the 2D case. Therefore, beside the much less computational cost, an important
advantage of the 2D simulations is that the role of long-range interactions in
irregular plastic yielding is isolated, so its effects can be better studied.

\begin{figure}[!ht]
\begin{center}
\vspace*{-0.2cm}
\includegraphics[angle=-90,width=7cm]{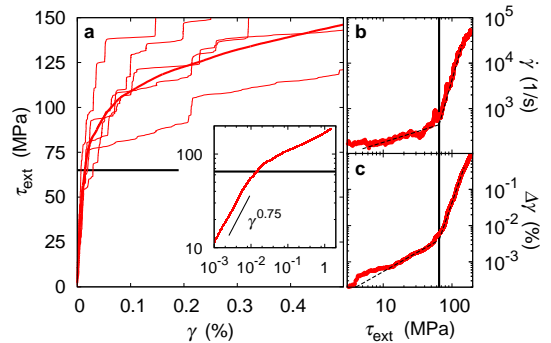}
\end{center}
\vspace*{-0.4cm}
\caption{Average properties of strain responses obtained by DDD simulations in tree dimensions.
(a) Stress ($\tau_\text{ext}$) versus plastic strain
($\gamma$) curves of individual simulations (thin lines). The thick line is the
stress-strain curve averaged over samples. The black straight line marks the
suggested critical yield stress. Inset shows the averaged curve on a log-log
plot. (b) Deformation rate versus external stress.
(c) Root mean square of the plastic strain fluctuations as
function of the external stress. \label{fig:2}}
\end{figure}

The plastic strain responses measured during individual stress-controlled 3D
simulations are seen in Fig.~\ref{fig:2}(a) (thin lines). The obtained stress
($\tau_\text{ext}$) vs.~plastic strain ($\gamma$) curves exhibit random steps,
just like the ones measured experimentally by microcrystal
deformation \cite{uchic_science, dimiduk_science}. The plateaus clearly indicate
strain avalanches, resulting in different patterns for different samples
excluding any practical definition of a threshold value. If, however, we average
over the more than 100 independent realizations of submicron samples, the
cavalcade of random staircases smoothens into a continuous curve, the thick line
of Fig.~\ref{fig:2}(a) Moreover, as seen in the inset of Fig.~\ref{fig:2}(a), there is a threshold
stress value $\tau_\text{c}\approx 65 \pm 10$ MPa marking the end of the pure
power region. The onset of the flow is even more evident in the deformation rate
$\dot{\gamma}$ vs.~external stress relation, plotted in Fig.~\ref{fig:2}(b), undergoing a
quite sharp transition at the same $\tau_\text{c}$. Furthermore, the root mean
square fluctuations over samples of plastic strain values at given stresses,
depicted in Fig.~\ref{fig:2}(c), sharply increase for $\tau_\text{ext} > \tau_\text{c}$.

\begin{figure}[!ht]
\vspace*{-0.2cm}
\begin{center}
\includegraphics[angle=-90,width=7cm]{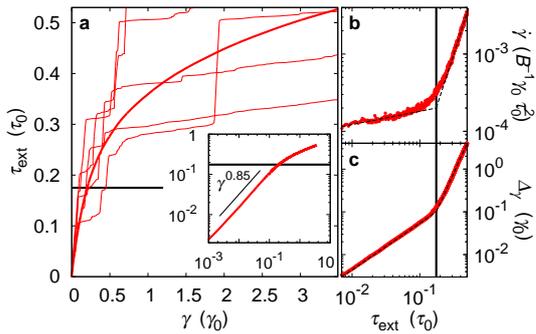}
\end{center}
\vspace*{-0.4cm}
\caption{Average properties of strain responses obtained by DDD simulations in
two dimensions. (a,b,c) Each panel is the counterpart of
Figs.~\ref{fig:2}(a-c), now obtained by 2D simulations. Here the quantities are in natural
units defined in the text. \label{fig:3}}
\end{figure}

We now turn to the detailed presentation of the 2D results. The simulations were
started out from different random distributions with zero net Burgers vector.
For each of the different 5000 realizations, after letting the system relax at
zero external stress, stress-controlled loading was applied. In contrast to 3D, in the 2D simulations all the material parameters can be scaled out
by introducing natural units as $v_0 = \sqrt{\rho} B^{-1} G b^2$ for velocity, $\gamma_0 = b \sqrt{\rho}$ for plastic strain and $\tau_0 =
Gb\sqrt{\rho}$ for stress, where $\rho$ is the total dislocation density and $G$
is a combination of elastic moduli \cite{bako_07}. Just like in three
dimensions, the individual curves are step-like (Fig.~\ref{fig:3}(a)), but again, a
relatively sharp critical transition point can be identified on the external
stress dependence of the average plastic strain (inset of Fig.~\ref{fig:3}(a)), on the
deformation rate $\dot{\gamma}$ (Fig.~\ref{fig:3}(b)) and also on the plastic strain
fluctuation (Fig.~\ref{fig:3}(c)). For 128 dislocations the critical stress obtained is
$\tau_\text{c}\approx (0.17\pm 0.02) \, \tau_0$. The outstanding similarity
between the strain responses of 2D and 3D systems observed in Figs.~\ref{fig:2} and \ref{fig:3}
suggests that the simplified 2D case, only containing the long-range
interactions between dislocations, is able to capture the general features of
submicron plastic flow. We can conclude that not only there is in the
statistical sense a smooth stress-strain curve for submicron sizes, but also
that several corroborating indicators show the existence of a quite sharply
defined threshold stress, presumably a material characteristics for a given
specimen size and initial dislocation density.

\begin{figure}[!ht]
\vspace*{-0.2cm}
\begin{center}
\includegraphics[angle=-90,width=7cm]{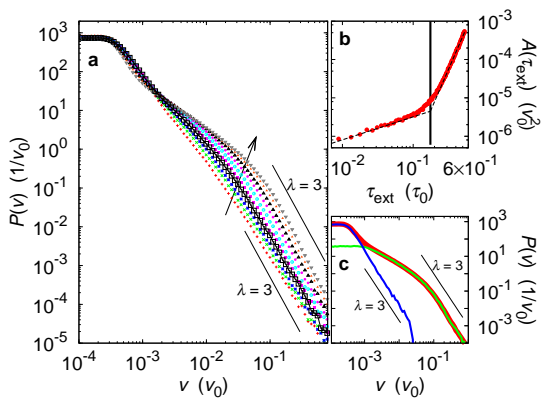}
\end{center}
\vspace*{-0.4cm}
\caption{(a) Dislocation velocity distributions at different
external stress levels. The arrow indicates the direction of increasing external
stresses. (b) The prefactor $A(\tau_\text{ext})$ (see equation
(1)) as a function of the external stress. (c) Above the
yielding transition the velocity spectrum (red) can be decomposed into the
contribution of avalanches (green) and that of the quiescent configurations
(blue). \label{fig:4}}
\end{figure}

More can be learned from the detailed analysis of the dislocation segment
velocity distribution $P(v)$. Since the \emph{in situ} direct measurement of the
velocities of large number of dislocations is rather difficult, there are no
experimental data available on them so far. Here we report about results
obtained by DDD simulations in 2D. In each individual simulation we determine
the distribution of the absolute velocities of the dislocation segments, at
different load levels  and take their averaged over 5000 statistically
equivalent realizations (see Fig.~\ref{fig:4}(a)). The first thing that needs to be
mentioned is the remarkable fact that the tails of the distributions exhibit a
scale-free
\begin{equation}
	P(v) \approx A v^{-\lambda}
\label{eqn:power_law_tail}
\end{equation}
power-law asymptotic decay with exponent $\lambda \approx 3 \pm 0.02$.

The theoretical explanation of the cubic decay can be obtained under some natural
assumptions about the correlation of dislocations.  Previously  it was  shown that the distribution of internal elastic stresses at a random location has a cubic tail \cite{groma_bako_98}, presently however we are interested in the stress (proportional to the velocity) at the positions of dislocations.  We give a simplified derivation first and then discuss its adaptation to our case.  The cumulant generating function $\Psi(q)$ of the stress distribution is given by
\begin{equation}
\begin{split}
 e^{\Psi(q)}&= \int \rmd\tau \, e^{iq\tau} \, P(\tau) = \left\langle
e^{iq\sum_{n=1}^N\tau(\Bm{r}_n-\Bm{r}_0)}\right\rangle ,\label{eq:mom-gen}
\end{split}
\end{equation}
where the average is taken over the normalized joint distribution of $N+1$
dislocations  $f(\BM{r}_0,\dots\BM{r}_N)$ and  $\tau(\BM{r})=r^{-1}\cos\varphi\,
\cos2\varphi$ is the stress field generated by a dislocation,  in appropriate units, using polar coordinates. The large-$\tau$ behavior is determined by small $q$'s,
so we keep the leading term in the Mayer cluster expansion
\begin{equation}
\begin{split}
 \Psi(q)  &\approx   N
\int \left( e^{iq\tau(\Bm{r})}-1\right)f(\BM{r}|\BM{0})\, \rmd^2r,
\end{split}
\end{equation}
where the conditional distribution $f(\BM{r}|\BM{0})$ appears.  For
$r<q\epsilon$, where $\epsilon$ is small but fixed, the phase factor oscillates
fast so the integral gives at most an order $q^2$ contribution, if $f$ is
nonsingular.  Setting the average of $\tau$ to zero, the leading term
comes from expanding in $q$ as
\begin{equation}
 \Psi(q) \approx - \frac N2 q^2\int\limits_{r>q\epsilon}
\tau^2(\BM{r})\, f(\BM{r}|\BM{0})\, \rmd^2r \approx - \frac{A'}{2} \, q^2\, \ln
q,
\label{eq:mom-gen2}
\end{equation}
where  $A'=N \int_0^{2\pi}\!\! \cos^2 \varphi\, \cos^2 2\varphi\, f(r\!\!=\!\!0, \varphi|\BM{0})\, \rmd\varphi\,$,  provided the small-distance limit of $f(r,\varphi|\BM{0})$  depends only on the angle.  Note that $1/f$ is of the order of  area, so $Nf$ is nonextensive.  As the final step in the derivation one straightforwardly shows that the asymptote $P(\tau)\approx A' \tau^{-3}$ yields by \eqref{eq:mom-gen} the $q$-dependence in \eqref{eq:mom-gen2}. We mention that the cubic decay implies a variance logarithmically diverging in the upper cutoff, corresponding to the logarithmic factor in \eqref{eq:mom-gen2}, which is a typical feature of X-ray line-profile tails \cite{groma_98}.

Now we need to refine the above picture by the following considerations.
Firstly, there are two types of dislocations with $\pm$ Burgers vectors, so
pair correlations between all types have to be introduced, as it was worked out in \cite{groma_bako_98}.  Secondly, we should realize that the
equilibrium correlation $f_\mathrm{eq}(\BM{r}|\BM{0})$ may diverge for small
distances \cite{groma_06}. However, since dislocations do not move, the stress distribution is the Dirac delta centered at the origin, so $\Psi_\mathrm{eq}(q)\equiv 0$.  Thus we can use in \eqref{eq:mom-gen2} the time dependent deviation from the equilibrium
correlation. A detailed analysis will be published elsewhere.  In conclusion, if the time-decaying term in the correlation produces a finite amplitude in \eqref{eq:mom-gen2}  then the stress and thus the velocity distribution must have a reciprocal cubic decay.

Returning to the discussion of the simulations, the prefactor $A$ in Eq.\
\eqref{eqn:power_law_tail} was found to depend on the momentary external stress
$\tau_\text{ext}$. According to Fig.~\ref{fig:4}(b), $A(\tau_\text{ext})$ also undergoes a
transition. Although the turn is not very sharp, the asymptotes apparently have
different slopes, whose intersection point again gives
$\tau_\text{c}$ of Fig.~\ref{fig:3}(a-c).
In a statistical sense, therefore, $\tau_\text{ext}>\tau_c$ corresponds to the flowing regime, so,  for a given size on the average, a dynamical transition takes place at $\tau_\text{c}$, which can be considered as a measure of the strength of the material.  It should be stressed, however, that in an individual sample considerable dislocation motion may occur below $\tau_\text{c}$. One can expect that, for samples with macroscopic sizes, $\tau_\text{c}$ goes over to the conventional yield stress.  Finally we note that according to our preliminary investigations the rate dependence seems to be weak.

The investigations on the velocity distributions have been repeated also in 3D. Although in this case only a smaller ensemble can be
afforded and the inherent numerical noise is much higher, there is a striking
similarity between the 3D and the 2D results discussed above. The tail of the
distribution is cubic, and its prefactor increases with the growing external
stress.  The emergence of the cubic decay indicates that its derivation in
the 2D single-slip case actually has a much broader validity. Furthermore, as
in 2D, the different distributions separate after a well-defined point and the
small-$v$ part is nearly unaffected by the external stress. We must conclude
that the observed velocity distribution and its evolution are of universal
nature.

\begin{figure}[!ht]
\begin{center}
\includegraphics[angle=-90,width=7cm]{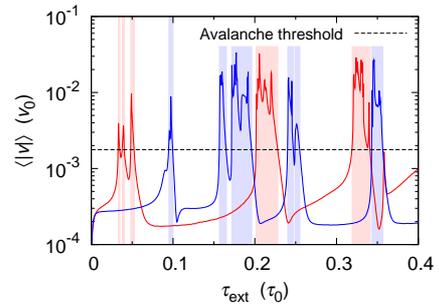}
\vspace*{-0.6cm}
\end{center}
\caption{Identification of avalanches in individual simulation runs. The average
of the absolute velocity fluctuates during a simulation. In the figure the two
curves correspond to two different realizations, but all simulation parameters
are the same. The dashed line denotes the threshold value used for avalanche
identification. The regions, where an individual run is in avalanche state are
marked by the corresponding light colors. \label{fig:5}}
\end{figure}

The next issue to be considered is whether there is any mark of the avalanche
activity on the velocity distribution of dislocations. In an individual
simulation run the system alternates between the quiescent and avalanche states.
This is manifested in surges in the average absolute dislocation velocity within
a sample as function of time, the heights identified as avalanches separated by
low activity periods (see Fig.~\ref{fig:5}).
These two states can be distinguished by
thresholding the average dislocation velocity in each separate run. At a given
external stress level it is instructive to calculate separately the averaged
velocity distributions corresponding to systems being in quiescent and avalanche
states. The two distributions are plotted in Fig.~\ref{fig:4}(c) with blue and green curves,
respectively. As seen, the tail of the total velocity distribution comes from
systems being in avalanche state. Samples in quiescent states contribute only to
the small-$v$ part of $P(v)$. The superposition of the two convex curves results
in the characteristic ``shoulder'' in the distribution functions plotted in
Fig.~\ref{fig:4}(a). These observations apply in three dimensions, too. In sum, the velocity
distribution exhibits universality in more than one respect. First, the
low-velocity part of the histogram is nearly independent of the external load. Second,
the exponent of the decay at large velocities appears to be constantly $3$ all
along the loading scenario. Third, the critical yield stress manifests itself by
an upturn in the amplitude of the power tail, due to increased avalanche weight.

While so far we analyzed peculiarities of the velocity distribution of
dislocations, Orowan's well known law \cite{hirth_lothe} states that the
plastic strain rate is proportional to the average dislocation velocity,
weighted by the Burgers vector. So, it is natural to ask what the contribution
is of the distribution tail obtained above, corresponding to avalanches, to the
average plastic rate. We found that about $80$\% of the plastic strain comes
from the region at and above the ``shoulder''.  At the same time $20$\% comes from the power-decaying part, significant for a tail. This implies that the fast dislocations forming avalanches play dominant role in the plastic response.

In summary we should emphasize that small scale dislocation simulations abound
from the past thirty years. On the other hand, experiments were available only
on macroscopic plasticity, in sizes never reached by simulations. In the present
paper the two approaches meet, now that experiments reached down to submicron
level, simulations become more faithful. Increased computer power made us
possible to describe ensembles never considered before, thus a statistical
analysis with new conclusions and definition of the yield stress in small scale
specimens could be reached.

\begin{acknowledgements}
The authors thank G.\ T.\ Zim\'anyi for illuminating discussions and comments. G.\ Gy.\ is indebted to M.\ Droz for hospitality at the University of Geneva.
Financial supports of the Hungarian Scientific Research Fund (OTKA) and by the
Swiss FNRS are acknowledged.
\end{acknowledgements}

\bibliography{submic}

\end{document}